
\catcode `\^^Z=9                   

\magnification=\magstep1

\vsize 9truein
\hsize 6.5truein
\voffset -.25truein
\hoffset -.25truein
\looseness=-1
\tolerance=1600
\linepenalty=100
\widowpenalty=4000
\clubpenalty=4000
\hbadness=10000
\vbadness=10000
\baselineskip=11pt         

\def\doublespaced{\baselineskip=1.6em \lineskip = 0.1em \lineskiplimit
= 0.1em}

\vsize 9truein
\hsize 6.5truein
\voffset -.25truein
\hoffset -.25truein
\looseness=-1
\tolerance=1600
\linepenalty=100
\widowpenalty=4000
\clubpenalty=4000
\hbadness=10000
\vbadness=10000
\baselineskip=11pt         

\def\doublespaced{\baselineskip=1.6em \lineskip = 0.1em \lineskiplimit
= 0.1em}

\def\disclaim{\nine This work was supported in part by the High Energy
Theory Physics
 program of the U.S. Department of Energy under Contract
No.~DE-AC02-76CH00016 and in part by NSF Grant No.~PHY-9017585.}

\def\v{\vskip 1truepc}

\def\b{\bigskip}

\def\lsim{\mathrel{\lower4pt\hbox{$\sim$}}
\hskip-10pt\raise1.6pt\hbox{$<$}\;}

\def\gsim{\mathrel{\lower4pt\hbox{$\sim$}}
\hskip-10pt\raise1.6pt\hbox{$>$}\;}

\def\frac#1#2{{#1\over#2}}

\def\chass{{\cal O}(\hat\alpha^2_s)}
\def\drho{\Delta\rho}
\def\drhof{(\Delta\rho)_f}
\def\hals{\hat\alpha_s}
\def\hmt{\hat m_t}
\def\ms{\overline{\rm MS}}
\def\cO{{\cal O}}
\def\PL{Phys.\ Lett.\ }
\def\NP{Nucl.\ Phys.\ }
\def\PR{Phys.\ Rev.\ }

\font\nine=cmr9
\font\mib=cmmib10
\vglue 1truein

\centerline{\bf QCD Corrections to $\mib
\Delta\rho$\footnote{*}{\disclaim}} \v\v

\centerline{A.\ Sirlin\footnote{**}{Permanent address: Dept.\ of
Physics, New York University, 4 Washington Place, New York, NY\ \
10003}} \b

\centerline{\it Department of Physics}
\centerline{\it Brookhaven National Laboratory}
\centerline{\it Upton, Long Island, NY\ \ 11973}
\v\v

\centerline{\bf Abstract}
\v

{\narrower \smallskip \noindent\nine We discuss a simple method to
evaluate the QCD corrections to $\Delta\rho$. It assumes that the
perturbative expansion in terms of $\ms$ parameters is meaningful and,
unlike other studies, exploits significant available information
concerning $\chass$ corrections. This approach leads to an enhancement
of $\sim26\%$ relative to the conventional evaluation. QCD corrections
to the $Z^0\to b\bar b$ amplitude are also considered. Implications for
electroweak physics are briefly discussed. \smallskip}
\v\v
\doublespaced

The $\cO(\alpha\alpha_s)$ corrections to the electroweak amplitudes
$\drho$ and $\Delta r$ have been the object of detailed studies in the
past [1]. These analyses have been recently extended to the
contributions proportional to $m^2_t$ in the $Z^0\to b\bar b$ amplitude
[2]. In a recent paper, Smith and Voloshin (S-V) have re-examined the
$\cO(\alpha\alpha_s)$ corrections to $\drho$ by a detailed investigation
of the relevant Feynman diagrams involving the $t$-$b$ isodoublet [3].
Because of the large mass of the top quark and the fact that $\drho$ is
evaluated at a much smaller scale, namely $q^2=0$, the conventional
wisdom is that the Feynman integrals are dominated by euclidean gluon
momenta $k\sim m_t$ [4]. In fact, in most of the published calculations,
$\hals$ is evaluated at $m_t$, with the cases $\hals(2m_t)$ and
$\hals(m_t/2)$ sometimes considered to estimate the theoretical error.
However, in the case of $\drho$, S-V find a significant sensitivity to
scales $\sim0.15 m_t$. Specifically, employing the V-scheme running
coupling [5], carrying out the integration over the euclidean gluon
momentum and absorbing the result in a rescaling of $\hals$, these
authors find:

$$ \drhof = x_t \left[ 1-\frac{2\hals(0.154 m_t)}{9\pi} (\pi^2+3)
\right], \eqno (1) $$

\noindent where the subscript $f$ means fermionic contribution,

$$ x_t \equiv \frac{3G_\mu m^2_t}{8\pi^2 \sqrt{2}} \eqno (2) $$

\noindent is the one-loop correction [6] and $m_t$ the on-shell mass.
Eq.~(1) is the usual perturbative result [1], except that the $\ms$
coupling $\hals$ is evaluated at $0.154 m_t$, rather than the scale
$m_t$ conventionally employed in these calculations. The authors of
Ref.~[3] point out that a complete three-loop $\cO(\alpha\alpha^2_s)$
calculation is needed to completely quantify this scale. However, as
things presently stand, the shift in the $\hals$ scale from $m_t$ to
$0.154 m_t$ is quite large, and implies a significant modification in
the $\cO(\alpha\alpha_s)$ corrections. For instance, for
$\hals(m_Z)=0.118$ and $m_t=200$ GeV (roughly the upper bound of the
current indirect determination of $m_t$), $\hals(0.154
m_t)/\hals(m_t)=1.34$, where we have employed a three-loop evaluation of
$\hals$. Thus, the S-V rescaling of $\hals$ leads to a $\approx34\%$
enhancement of the usual QCD contribution to $\drhof$. Furthermore, as
the corrections of $\cO((\hals/\pi)^2)$ involved in this rescaling have
large coefficients $\approx-20.5$, one wonders whether the neglected
$\cO(\hals^2)$ corrections may also be sizable.

In this communication we discuss a simple method to evaluate the QCD
contributions to $\drhof$ which, in contrast with other studies,
exploits significant available information concerning $\cO(\hals^2)$
corrections. As we will see, it also assumes that the perturbative
expansion of $\drhof$ in terms of $\ms$ parameters is meaningful. We
start with the observation, made by S-V, that the sensitivity to
relatively low mass scales exhibited in Eq.~(1) arises from the use of
the on-shell mass $m_t$ in Eqs.~(1) and (2). In fact, these authors
traced such dependence to the contribution to $\drhof$ of the mass
counterterm associated with $m_t$. In order to circumvent this problem,
we propose to first express $\drhof$ in terms of $\hmt(m_t)$, the
running $\ms$ mass evaluated at the $m_t$ scale. One readily obtains

$$ \drhof = \frac{3G_\mu\hmt^2}{8\pi^2\sqrt{2}} \left[ 1-\frac{2}{9}\;
\frac{\hals}{\pi} (\pi^2-9) + C\left( \frac{\hals}{\pi} \right)^2 +
\cdots \right], \eqno (3) $$

\noindent where $\hmt$ is henceforth an abbreviation for $\hmt(m_t)$,
$\hals$ is also evaluated at $m_t$, $C$ is the coefficient of the
$\cO((\hals/\pi)^2)$ contributions which have not been calculated so
far, and the ellipses represent higher order terms. As the answer is
expressed in terms of $\hmt$, the corrections in Eq.~(3) should not be
sensitive to the relatively low mass scales found by S-V and, in fact,
the dominant contributions are expected to involve euclidean momenta
$k\gsim m_t$. Furthermore, we note that the coefficient of the
$\cO(\hals/\pi)$ term is small, namely $\approx-0.1932$. We now recall
that the relation between $\hmt$ and $m_t$ is accurately known [7]:

$$\frac{m_t}{\hmt} = 1+\frac{4\hals}{3\pi} + \left[ 16.11 - 1.04
\sum^5_{i=1} \left(1-\frac{m_i}{m_t} \right) \right]
\left(\frac{\hals}{\pi} \right)^2 +\cdots, \eqno (4) $$

\noindent where $\hals$ is again evaluated at $m_t$. For 130 GeV${}\lsim
m_t \lsim 200$ GeV, we approximate the small $m_i$-dependent term within
the square brackets by 0.04 and rewrite Eq.~(4) as

$$\frac{m_t}{\hmt} = 1+ \frac{4}{3}\; \frac{\hals}{\pi} + [16.15 - 1.04
n_f] \left( \frac{\hals}{\pi} \right)^2 \cdots, \eqno (5) $$

\noindent where $n_f=5$ is the number of light quark flavors.
Numerically, the coefficient of the $\cO((\hals/\pi)^2)$ term is quite
large, namely 10.95, which may partly reflect the sensitivity to the
relatively small mass scales discussed before. In order to improve the
convergence of the expansion, we apply the BLM method [5]. Writing
$n_f=33/2-3\beta_0/2$, where $\beta_0$ is the coefficient of the
one-loop $\beta$ function for $\alpha_s$, one absorbs the $\beta_0$ term
in a shift in the $\hals$ scale, while $-1.04\times33/2$ is combined with
16.15. Eq.~(5) becomes then

$$\frac{m_t}{\hmt} = 1+\frac{4}{3}\; \frac{\hals}{\pi} (Q^\ast) \left[
1-0.758 \frac{\hals(Q^\ast)}{\pi} \right] + \cdots, \eqno (6) $$

\noindent where $Q^\ast=0.0963 m_t$. This procedure absorbs vacuum
polarization contributions in the coupling constants. We see that
$Q^\ast \ll m_t$, a frequent feature of the BLM method, but at the same
time the convergence of the series is much improved. For example, for
$m_t=200$ GeV and $\hals(m_Z)=0.118$, we have $\hals(m_t)=0.1055$ and
$\hals(Q^\ast)=0.1546$, so that Eq.~(5) gives $1+0.0448+0.0123 + \cdots
= 1.0571 + \cdots$, while Eq.~(6) leads to $1+0.0656 - 0.0024 + \cdots =
1.0632 + \cdots$. We also note that the two-loop result from Eq.~(6) is
larger by $6\times10^{-3}$. As Eq.~(6) contains a very small next to
leading term, it gives essentially the same result as the method of
fastest apparent convergence (FAC) [8]. The latter would absorb the complete
$\cO(\hals^2)$ contribution in Eq.~(5) into a rescaling of the
$\cO(\hals)$ term. In the following, we employ Eq.~(6).

Combining Eqs.~(3) and (5), we have

$$\drhof = x_t \frac{\left[ 1-\frac{2}{9}\;\frac{\hals}{\pi} (\pi^2-9) +
C \left(\frac{\hals}{\pi} \right)^2\right]}{(m_t/\hmt)^2}, \eqno (7) $$

\noindent where $\hals$ in the numerator is evaluated at $m_t$ and
$m_t/\hmt$ is given by Eq.~(6). For a very precise evaluation of
Eq.~(7), the determination of the constant $C$ would be needed and this
is not available at present. However, if we make the plausible
assumption that the $\cO(\hals^2)$ term in Eq.~(3) and the numerator of
Eq.~(7) is not larger than the corresponding $\cO(\hals)$ contribution,
it is clear that the bulk of the QCD correction resides in the
denominator of Eq.~(7). In fact, this factor contains the dependence on
the relatively small mass scales discussed above and, moreover, it
clearly absorbs most of the $\cO(\hals)$ contribution. It is also
interesting to note that if one expands Eqs.~(1) and (7) in powers of
$\hals(m_t)$ up to terms of $\cO(\hals^2)$, one can match the
second-order coefficients by choosing $C=-4.46$. This means that the
denominator of Eq.~(7) absorbs most of the $\cO(\hals^2)$ contribution
of Eq.~(1) ($-16.1(\hals/\pi)^2$ out of $-20.5(\hals/\pi)^2$ to be
precise). The important theoretical point is that this large
contribution, $(m_t/\hmt)^2$, has been evaluated with considerable
accuracy (Eqs.~(5) and (6)). On the basis of these arguments, we propose
to calculate $\drhof$ using Eq.~(7) with $\hals$ evaluated at $m_t$ in
the numerator, and $m_t/\hmt$ given by Eq.~(6). As the constant $C$ has
not been determined, we treat the last term in the numerator of Eq.~(7)
as a measure of the theoretical error associated with unaccounted higher
order corrections. Assuming that the perturbative expansion in Eq.~(3)
is meaningful, we fix $|C|$ so that the $\cO(\hals^2)$ term in that
equation is not larger in absolute value than the $\cO(\hals)$
contribution. This leads to $C\approx\pm6$. With this choice, the
coefficient of $(\hals/\pi)^2$ in Eq.~(3) and the numerator of Eq.~(7)
may be as large as $6/0.1932\approx31$ times the cofactor of
$\hals/\pi$. By way of comparison, in the case of
$\Gamma(\Upsilon\to\hbox{hadrons})/\Gamma(\Upsilon\to\mu^+\mu^-)$, which is
often cited as an illustration of large higher order coefficients, the
cofactor of the next to leading term in an expansion in powers of
$\hals(M_\Upsilon)/\pi$ is 9.1 relative to the leading contribution; in
the case of $\Gamma(\eta_c\to\hbox{hadrons})/\Gamma(\eta_c\to
\gamma\gamma)$, an expansion in powers of $\hals(M_{\eta_c})/\pi$ shows
a 14.5 relative coefficient [5]. We may also attempt to estimate the
vacuum polarization contribution to $C$ by combining Eqs.~(1,5,7): one
finds $C_{\rm vac. pol.}\approx0.44 \beta_0$. As in this case the
relevant scale is $k>m_t$, it is natural to consider six active quark
flavors so that $C_{\rm vac. pol.}\sim3.1$. (These ``vacuum polarization''
effects include gluonic contributions associated with the running of
$\hals$). Another way of estimating $C_{\rm vac. pol.}$
 is to apply the BLM method and absorb this contribution in a
rescaling of the $\cO(\hals)$ term in Eq.~(3). Evaluating accurately the
rescaled term one finds that, for $m_t=200$ GeV, the result is
equivalent to using an effective value $C_{\rm vac. pol.}\approx2.2$ in
Eq.~(3). We note that both estimates are positive, which is consistent
with the notion that the dominant contributions to Eq.~(3) involve
$k>m_t$. Yet another approach to estimate $C$ is to apply the principle
of minimum sensitivity (PMS) [9]. One expresses Eq.~(3) in terms of
$\hat m^2_t(\mu)$ so that $\drhof$ is proportional to $\hat m^2_t(\mu)
\{1+ (\hals(\mu)/\pi) [4\ell n (\mu/m_t) - (2/9) (\pi^2-9)]\}$.
Evaluating this quantity at the stationary point $\mu^\ast$, which
according to the PMS is the optimal value, we find that, for $m_t=200$
GeV, the result is equivalent to $C\approx2.7$ in Eq.~(3). Thus, we see
that these estimates of $C$ are roughly consistent and all comfortably
lie within our error range. On the basis of the various arguments given
above, it appears that our choice $C=\pm6$ in the error analysis is
quite conservative. Another possible source of theoretical uncertainty
arises from the $\cO((\hals(Q^\ast)/\pi)^3)$ contributions to Eq.~(6).
If, following the pattern of Eq.~(6), the corresponding coefficient is
$\sim1$, such terms would give very small effects${}\sim
\pm2\times10^{-4}$. If the coefficient is $\sim10$, their effect would
be $\sim\pm2\times10^{-3}$, three times smaller than the estimated error
we have included. Using then $\hals=\hals(m_t)=0.1055$,
$\hals(Q^\ast)=0.1546$ (values corresponding to $m_t=200$ GeV and
$\hals(m_Z)=0.118$), and $C=\pm6$, Eq.~(7) gives, for the QCD correction
to $\drhof$:

$$ \drhof -x_t = -x_t[0.121\pm0.006]. \eqno (8) $$

\noindent  Eq.~(8) is to be compared with the conventional result
$\drhof-x_t =-0.0960x_t$ (Eq.~(1) with $\hals(m_t)$) and the S-V
proposal $\drhof-x_t=-0.129x_t$ (Eq.~(1) with $\hals(0.154m_t)$). Thus,
Eq.~(8) leads to a 26\% enhancement of the QCD correction over the
conventional answer, and a 5\% error estimate, relative to the
calculated QCD effect, due to the higher order corrections. It is
somewhat smaller than the S-V result, but roughly consistent with it if
one takes into account the estimated uncertainty. A more complete
discussion of the error must include the effect introduced in Eq.~(8) by
the experimental uncertainty in $\hals(m_Z)$. For $\delta\hals(m_Z) =
\pm0.006$, we find an additional error $\pm0.007$ in Eq.~(8). Combining
the two errors in quadrature, the overall uncertainty in Eq.~(8) becomes
$\pm0.009$, or about 8\% (if added linearly the total error would be
$\pm0.013$ or 11\%).

For $m_t=200$ GeV, the difference between Eq.~(8) and the conventional
evaluation leads to an additional contribution $-3.1\times10^{-4}$ to
$\drhof$ and $+1.1\times10^{-3}$ to $\Delta r$ [10]. The latter
corresponds to a decrease $\approx 19$ MeV in the predicted value of
$m_W$ for given $m_t$, or an increase $\Delta m_t=+2.8$ GeV for given
$m_W$. On the other hand, the S-V proposal (Eq.~(1)) implies somewhat
larger shifts, namely $\Delta m_W=-25$ MeV and $\Delta m_t=3.7$
GeV\null.

An analogous argument can be applied to the term proportional to $m^2_t$
in the $Z^0\to b\bar b$ amplitude. Including the $\cO(\alpha\alpha_s)$
correction, it is of the form [2]

$$m^2_tK^{(v)}_b = m^2_t [ 1-(\pi^2-3) \hals/3\pi+ \cdots], \eqno (9) $$

\noindent where the superscript $v$ means ``vertex'' and the subscript
$b$ refers to $Z^0\to b \bar b$. In terms of $\hmt^2$, this becomes
$\hmt^2 [ 1+(11-\pi^2) \hals/3\pi + D (\hals/\pi)^2 + \cdots ]$. Thus we
see that, once more, the coefficient of the $\cO(\hals)$ term is
significantly reduced. The latter expression can be rewritten as

$$m^2_tK^{(v)}_b = m^2_t \frac{\left[ 1+(11-\pi^2) \frac{\hals}{3\pi} +
D(\hals/\pi)^2 \right] }{(m_t/\hmt)^2}, \eqno (10) $$

\noindent where again $\hals$ in the numerator is evaluated at $m_t$ and
$m_t/\hmt$ in the denominator is given by Eq.~(6). For $m_t=200$ GeV, we
choose $D=\pm11$ (which corresponds to the situation in which the
$\cO(\hals^2)$ and $\cO(\hals)$ terms in the numerator are equal in
magnitude), and obtain, for the QCD correction,
$m^2_t(K^{(v)}_b-1)=-m^2_t[0.104\pm0.011]$. This is to be compared with
the conventional result $-m^2_t0.0769$. So, in this case we find a 35\%
enhancement over the conventional answer and an 11\% error estimate,
relative to the calculated QCD effect, due to the higher order
corrections.

It should be pointed out that the use of the on-shell mass in Eq.~(2),
which is the usual choice, implicitly assumes that this is the relevant
parameter to discuss top physics: its production, decay and the
measurement of its mass [11]. For instance, if the running mass $\hat
m_t$ were found to be an appropriate parameter to analyze certain
aspects of top physics, one could directly use Eq.~(3). The idea of
expressing $\drhof$ in terms of $\hmt^2$, with or without additional
$\cO(\hals)$ corrections, and to employ Eq.~(5) up to $\cO(\hals/\pi)$
or higher to relate the two masses, has occurred to a number of authors
[11,12]. Our strategy has been to separate the QCD corrections to
$\drhof$ into two parts: 1) The factor $(m_t/\hmt)^2$, where the
corrections are large but known through $\cO(\hals^2)$. We have further
shown how the convergence of this series can be significantly improved
by applying the BLM method, and noted that essentially the same result
is obtained in the FAC approach. Thus, the present framework employs
significant information concerning $\cO(\hals^2)$ contributions. 2) The
second part involves the corrections to $\drhof$, when expressed in
terms of $\hmt^2$. Here the $\cO(\hals)$ contribution is small, while
the $\cO(\hals^2)$ term has not been computed so far. By assuming that
the latter is not larger than the former, so that perturbation theory
may be viable, the magnitude of the $\cO(\hals^2)$ term in that
expansion is bounded. We have also noted that this procedure allows for
the possibility that the coefficient of the  $(\hals/\pi)^2$ term may be
as large as $\approx 31$ times the cofactor of the $\hals/\pi$
contribution. In spite of this latitude, we obtain a rather small
estimate for the theoretical error of the overall perturbative QCD
correction. Analogous comments apply to the analysis of the QCD
corrections to the $m^2_t$ terms in the $Z^0\to b \bar b$ amplitude. Our
argument could go wrong in two ways: {\it i}) If the perturbative
expansion in Eq.~(3) and the analogous one for the $Z^0\to b\bar b$
amplitude fail. This would be the case if the coefficients $C$ and $D$
are significantly larger than the values $|C|=6$ and $|D|=11$ we
employed in the argument, or if, for special reasons, terms of nominal
$\cO(\hals^3)$ or higher are sufficiently large [13]. The elucidation of
these issues lies outside the scope of this paper. Here we have adopted
the point of view that the perturbative expansion in Eq.~(3) and the
corresponding one for the $Z^0\to b\bar b$ amplitude are meaningful,
which seems very reasonable in view of the large mass scales we are
considering. This point of view is supported by well-known theoretical
arguments invoked in the derivation of QCD sum rules for heavy quarks
[4,14]. If, in the future, additional significant contributions to
Eq.~(3) can be unambiguously shown to exist, they would have to be added
to the perturbative effects we have discussed. {\it ii}) If the
coefficient of the $\cO((\alpha_s(Q^\ast))^3)$ term in Eq.~(6) is
considerably larger than 10. In this regard, we point out that the nice
convergence pattern of Eq.~(6) is not suggestive of an explosive
behavior in the $\cO((\hals(Q^\ast))^3)$ contributions. Nonetheless, it
is clear that these studies will be greatly strengthened when complete
$\cO(\hat\alpha\hals^2)$ calculations become available and the constants
$C$ and $D$ are determined. They will also greatly benefit from the
analysis of $\cO((\hals(Q^\ast))^3)$ corrections to Eq.~(6). Meanwhile,
Eqs.~(6), (7), and (10) provide, in our view, a suitable perturbative
framework to evaluate the QCD corrections to $\drhof$ and $K^{(v)}_b$.
\v

\noindent {\bf Acknowledgements}. The author is indebted to P. Gambino,
B.A. Kniehl, W.J. Marciano, I. Papavassiliou, M. Schaden and A. Soni for
very useful discussions.
\v\v

\centerline{\bf REFERENCES}
\v
\parskip 11pt
\item{1.} A. Djouadi and C. Verzegnassi, \PL B{\bf195}, 265 (1987); A.
Djouadi, Nuovo Cimento A{\bf100}, 357 (1988); T.H. Chang, K.J.F.
Gaemers, and W.L. van Neerven, \NP B{\bf202}, 407 (1982); B.A. Kniehl,
J.H. K\"uhn, and R.G. Stuart, \PL B{\bf214}, 621 (1988); B.A. Kniehl,
\NP B{\bf347}, 86 (1990); F. Halzen and B.A. Kniehl , \NP B{\bf353}, 567
(1991); B.A. Kniehl and A. Sirlin, \NP B{\bf 371}, 141 (1992).

\item{2.} J. Fleischer, F. Jegerlehner, P. R\c aczka, and O.V. Tarasov,
\PL B{\bf293}, 437 (1992); G. Degrassi, \NP B{\bf407}, 271 (1993); G.
Buchalla and A.J. Buras, \NP B{\bf398}, 285 (1993).

\item{3.} B.H. Smith and M.B. Voloshin, UMN-TH-1241/94, TPI-MINN-94/5-T.

\item{4.} M.A. Shifman, A.I. Vainshtein, and V.I. Zakharov, \NP
B{\bf147}, 385 (1979).

\item{5.} S.J. Brodsky, G.P. Lepage, and P.B. Mackenzie, \PR D{\bf28},
228 (1983); S.J. Brodsky and H.J. Lu, SLAC-PUB-6389 (1993).

\item{6.} M. Veltman, \NP B{\bf 123}, 89 (1977); M.S. Chanowitz, M.A.
Furman, and I. Hinchliffe, \PL B{\bf78}, 285 (1978).

\item{7.} N. Gray, D.J. Broadhurst, W. Grafe, and K. Schilcher, Z.\
Phys.\ C{\bf48}, 673 (1990).

\item{8.} G. Grunberg, \PL B{\bf95}, 70 (1980) and B{\bf110}, 501
(1982); \PR D{\bf29}, 2315 (1984).

\item{9.} P.M. Stevenson, \PL B{\bf100}, 61 (1981); \PR D{\bf23}, 2916
(1981); \NP B{\bf203}, 472 (1982) and B{\bf231}, 65 (1984).

\item{10.} A. Sirlin, \PR D{\bf22}, 971 (1980) and D{\bf29}, 89 (1984).

\item{11.} See, for example, S. Fanchiotti, B. Kniehl, and A. Sirlin,
\PR D{\bf48}, 307 (1993), Section III.

\item{12.} F. Jegerlehner, in ``Progress in Particle and Nuclear
Physics'', edited by A. Faessler (Pergamon, Oxford, 1991); W.J.
Marciano, BNL-49236 (1993).

\item{13.} B.A. Kniehl, J.H. K\"uhn, and R.G. Stuart, in ``Polarization
at LEP'', CERN 88-06, p.~158 (1988) and paper cited in Ref.~[1]. For a
recent implementation of the dispersive approach pioneered in those
papers, see B.A. Kniehl and A. Sirlin, \PR D{\bf47}, 883 (1993).

\item{14.} V.A. Novikov, L.B. Okun, M.A. Shifman, A.I. Vainshtein, M.B.
Voloshin, and V.I. Zakharov, Phys.\ Rep.\ {\bf41C}, 1 (1978).

\bye